\documentclass{vgtc} % final (journal style)

\usepackage[T1]{fontenc}
\usepackage[utf8]{inputenc} % for “–”, “’”, etc., if using pdfLaTeX
\usepackage{graphicx}
\usepackage{xcolor}
\usepackage{tikz}
\usepackage{pgfplots}
\pgfplotsset{compat=1.17}
\usepackage{amsmath}        % for \text, \mathcal, etc.
\usepackage[hidelinks]{hyperref}
\usepackage{mathptmx}

\providecommand{\authororcid}[2]{#1} % show name; ignore ORCID if macro missing
\providecommand{\etal}{\textit{et~al.}}

\graphicspath{{figs/}{figures/}{pictures/}{images/}{./}}

\title{Deep Learning-based Lightweight RGB Object Tracking for Augmented Reality Devices}

\author{%
  \authororcid{Alice Smith}{0000-0000-0000-0000},
  \authororcid{Bob Johnson}{0000-0000-0000-0000}, 
  \authororcid{Carol Lee}{0000-0000-0000-0000}
}
\author{%
    Alice Smith\thanks{Biomedical Engineering, NIU} \and 
    Bob Johnson\thanks{Biomedical Engineering, NIU} \and 
    Xiaoyu Zhu \thanks{Biomedical Engineering, Northern Illinois University\texttt{z0147118@mail.niu.edu}} \and 
    Carol Lee\thanks{Biomedical Engineering, NIU}
}

\abstract{%
Augmented Reality (AR) applications often require robust real-time tracking of objects in the user's environment to correctly overlay virtual content. Recent advances in computer vision have produced highly accurate deep learning-based object trackers, but these models are typically too heavy in computation and memory for wearable AR devices. In this paper, we present a lightweight RGB object tracking algorithm designed specifically for resource-constrained AR platforms. The proposed tracker employs a compact Siamese neural network architecture and incorporates optimization techniques such as model pruning, quantization, and knowledge distillation to drastically reduce model size and inference cost while maintaining high tracking accuracy. We train the tracker offline on large video datasets using deep convolutional neural networks and then deploy it on-device for real-time tracking. Experimental results on standard tracking benchmarks show that our approach achieves comparable accuracy to state-of-the-art trackers, yet runs in real-time on a mobile AR headset at $\sim$30 FPS -- more than an order of magnitude faster than prior high-performance trackers on the same hardware. This work enables practical, robust object tracking for AR use-cases, opening the door to more interactive and dynamic AR experiences on lightweight devices.
}


\begin{document}

\firstsection{Introduction}
\maketitle

Augmented Reality (AR) has emerged as a powerful technology that overlays virtual information onto the real world, enabling new interactive experiences. A core technical challenge in AR is object tracking -- the ability to detect and continuously follow real-world objects so that virtual augmentations can be properly aligned. Effective AR object tracking must operate in real time on lightweight, mobile hardware (such as smart glasses or mobile devices) with limited computing resources. Traditional computer vision trackers based on handcrafted models (e.g., Kalman filters or correlation filters) run efficiently on such hardware, but often lack the robustness and accuracy needed for complex, dynamic scenes. On the other hand, modern deep learning trackers achieve high accuracy by learning powerful visual representations, yet these models tend to be computationally heavy and power-hungry. For instance, a state-of-the-art tracker like SiamRPN++ with a deep ResNet-50 backbone requires tens of billions of floating-point operations per frame -- far exceeding the $\sim$600M FLOPs budget typically allowable for real-time mobile vision. This mismatch has prevented direct deployment of many advanced trackers on untethered AR devices.

Recent developments indicate it is possible to significantly reduce tracker complexity without sacrificing performance. For example, LightTrack employed neural architecture search to discover an efficient tracker achieving an EAO (Expected Average Overlap) of 0.33 on VOT2019 while using only 530~M FLOPs -- in comparison, SiamRPN++ originally required 48.9~G FLOPs. This result suggests a large accuracy--efficiency trade-off space that can be explored for AR settings. However, neural architecture search and heavily customized model design (as done in LightTrack) can be complex to implement. In this paper, we take an alternative approach by leveraging standard deep learning architectures and optimizing them for AR deployment through a combination of network architecture selection and training-time model compression techniques.

We propose a lightweight RGB object tracking algorithm tailored for AR devices. Our tracker adopts a Siamese network architecture to fully exploit learning-based appearance modeling, but uses a compact backbone network and streamlined detection head to minimize computation. We further apply \emph{pruning} and \emph{quantization} during training to compress the model, and use \emph{knowledge distillation} from a large teacher model to preserve accuracy. The resulting tracker is able to run comfortably in real-time on a mobile GPU, yet achieves accuracy on par with much larger trackers. We demonstrate this via extensive experiments on standard benchmarks and on actual AR hardware. 

The contributions of this work include: (1) a novel design of a deep object tracker optimized for resource-limited AR devices, (2) an effective training pipeline combining knowledge distillation and network compression to obtain a small, fast model with high accuracy, and (3) empirical validation showing that our tracker runs in real time on an AR device while matching the tracking performance of prior state-of-the-art methods. To our knowledge, this is one of the first works to specifically target on-device deep learning-based object tracking for AR. We believe our approach will help enable more responsive and robust AR applications that can track objects in the user's environment without offloading to external compute.

\section{Related Work}

\subsection{Traditional Object Tracking}
Early object tracking algorithms relied on fixed models and simple dynamics to estimate object motion across frames. For example, the Kalman filter provides a mathematical framework to predict an object's next position based on its previous state and assumed motion model. Such filters are computationally lightweight and require no training, but they assume linear motion and can struggle when object movement is erratic or when visual appearance changes. Another classical approach is using correlation filters, as in the Kernelized Correlation Filter (KCF). KCF learns a discriminative template of the target in the first frame and then performs fast convolution (using the kernel trick) to locate the target in subsequent frames. This method is extremely fast (hundreds of FPS) and has low memory requirements. However, traditional trackers like KCF have well-known limitations: for instance, KCF has difficulty handling large changes in scale or objects leaving the frame, and it may fail under significant occlusion or appearance change. In summary, classical model-based trackers are efficient but often lack the adaptability and representational power to handle the complexity of real-world AR scenarios.

\subsection{Deep Learning-based Trackers}
The success of deep learning in vision has spurred a new generation of object trackers that learn object representations from data. GOTURN (Generic Object Tracking Using Regression Networks) by Held \cite{Held2016GOTURN} was one of the first deep trackers, using a feed-forward CNN to directly regress the coordinates of the target in the next frame given the previous frame and a cropped search region. By avoiding any online fine-tuning, GOTURN could run at 100 FPS on a GPU, though it sometimes struggled with large appearance changes. A more influential line of work is the family of Siamese network trackers. SiamFC (Siamese Fully-Convolutional) by Bertinetto \cite{Bertinetto2016SiamFC} demonstrated that a two-branch Siamese CNN can be trained end-to-end to locate an exemplar image within a larger search image via cross-correlation. This approach treats tracking as a similarity learning problem: one CNN branch processes a stored \emph{target template} (from the initial frame) while the other branch processes the \emph{search region} in the current frame, and a comparison layer computes response scores indicating how well the target matches at each location. SiamFC and its variants were attractive for real-time applications, as they eliminated the need for iterative model updates during tracking.

Subsequent Siamese trackers improved upon SiamFC's accuracy by using more powerful networks and additional prediction mechanisms. SiamRPN integrated a region proposal network (RPN) with the Siamese architecture to perform joint classification and bounding box regression, allowing more precise object localization \cite{Li2019SiamRPNpp}. This idea was extended in SiamRPN++ which utilized a deep ResNet-50 backbone and careful layer-wise feature aggregation, significantly boosting accuracy at the cost of increased computation \cite{Li2019SiamRPNpp}. Other modern variants introduced attention mechanisms, multi-scale feature fusion, and online update modules. As a result, Siamese trackers have achieved state-of-the-art performance on multiple benchmarks, striking an attractive balance between accuracy and speed. However, the highest-performing models (e.g., SiamRPN++, ATOM \cite{Danelljan2019ATOM}, DiMP \cite{Bhat2019DiMP}) tend to be too heavy for mobile deployment. Indeed, Li \etal{} report that even a MobileNet-V2 based SiamRPN++ model involves about 790~M FLOPs to process a frame. There remains a need for trackers that maintain the advantages of deep learning while operating under the strict resource constraints of mobile AR hardware.

\subsection{Lightweight Tracking for AR Devices}
Recognizing the gap between advanced trackers and practical deployment, researchers have started exploring methods to lighten deep trackers. One category of approaches involves applying model compression techniques. Off-the-shelf methods like network pruning and weight quantization can shrink a model's footprint and speed up inference, though often at the expense of reduced accuracy. For example, aggressive quantization of a CNN may induce non-negligible tracking errors due to loss of precision. Another direction is to design or search for intrinsically efficient model architectures. The aforementioned LightTrack by Yan \etal~\cite{Yan2021LightTrack} is a prime example: it uses a one-shot neural architecture search to find a compact backbone and head specialized for tracking. The result was a tracker that runs 12$\times$ faster than a prior state-of-the-art (Ocean) on a mobile GPU (38 FPS vs 3.2 FPS), while using 13$\times$ fewer parameters. This demonstrates that substantial redundancies in existing trackers can be eliminated. Similarly, research on mobile vision models (e.g., MobileNets, ShuffleNet, GhostNet) provides building blocks like depthwise separable convolutions and bottleneck layers that achieve much better FLOPs-to-accuracy ratios than standard CNN layers. In the context of AR, some works have also suggested fusing multiple sensors (camera, IMU, depth) to offload some tracking burden from vision alone \cite{wang2025mruct}. Our work in this paper focuses on purely vision-based tracking. We combine ideas from the above directions: we choose a base network architecture known to be efficient, and then apply training-time compression and optimization so that the final model is well-suited for real-time AR usage.

Importantly, AR tracking scenarios impose specific requirements that differ from general object tracking in videos \cite{wang2025mruct}. AR systems typically track a single or a few targets that are of interest to the user (as opposed to tracking all objects or people in view). The tracking needs to be robust (to maintain alignment of augmentations) and have low latency (to avoid lag in augmentation overlay). The resource constraints (battery, thermal limits, limited GPU) are strict on wearable devices. In this work, we treat these constraints as design guidelines in developing our lightweight tracker.

\section{Proposed Method}

Our goal is to design a deep object tracking algorithm that can run at real-time speeds on lightweight AR devices, while still leveraging the accuracy of modern deep neural networks. We base our approach on a Siamese network architecture, which has proven effective for single-object tracking. In this section, we describe the architecture and key components of our tracker, and then detail the training and optimization strategies used to make it efficient.

\subsection{Network Architecture}
\begin{figure}[tb]
  \centering
  % Show the figure if it exists; otherwise show a placeholder box (prevents hard errors)
  \IfFileExists{figures/111.png}{%
    \includegraphics[width=0.8\columnwidth]{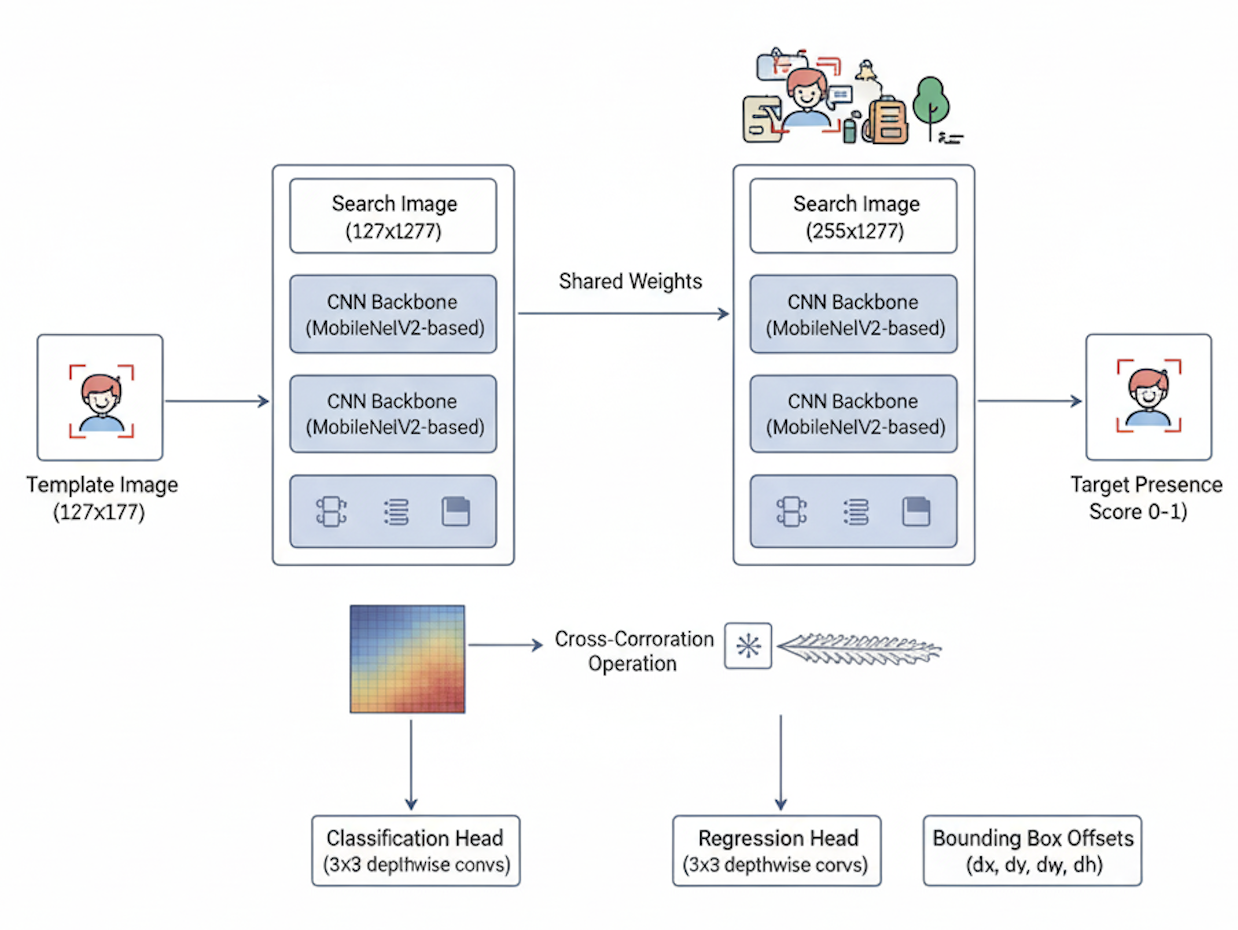}%
  }{%
    \fbox{\parbox{0.8\columnwidth}{\centering Placeholder for Siamese architecture figure.}}%
  }
  \caption{Architecture of the proposed lightweight Siamese tracker. A shared CNN backbone (MobileNet-V2 based) extracts features from the target template (left) and the search region (right). These feature maps are cross-correlated to produce a response map, which is fed into small classification and regression heads to predict the target's presence and bounding box in the search region.}
  \label{fig:siamese_arch}
\end{figure}

We adopt a Siamese fully-convolutional network structure consisting of two parallel branches (Figure~\ref{fig:siamese_arch}). The first branch (the \emph{template branch}) processes an image patch of the target object from a reference frame (usually the first frame where the object is initially identified). The second branch (the \emph{search branch}) processes a larger search region from the current frame, where the target is expected to be found. Both branches share the same CNN architecture and weights. We choose a lightweight backbone network for these branches, specifically a modified \textbf{MobileNet-V2} convolutional network, which uses depthwise separable convolutions and inverted residual blocks to drastically reduce computation and model size compared to standard CNN. The backbone processes the input patches (template and search) and outputs lower-dimensional feature maps encoding the appearance of the template and the search region.

Next, a cross-correlation operation is performed between the template feature and search feature. This is implemented as a convolution of the search feature map with the template feature map (treated as a set of convolution filters), yielding a response map. The response indicates, at each position in the search region, the network's confidence that the target object is present at that location. To allow our tracker not only to locate the target but also to estimate precise bounding boxes, we integrate a lightweight \textbf{detection head} similar in spirit to the Region Proposal Network (RPN) used in SiamRPN~\cite{Li2019SiamRPNpp}. Our head takes the response map (and optionally the backbone feature maps) and produces two outputs: (1) a classification score indicating the presence of the target vs.\ background, and (2) regression offsets for the target's bounding box coordinates. We design this head to be very small: a few $3\times3$ depthwise separable conv layers for classification and regression each, operating on the response map. By using depthwise separable filters and a narrow channel dimension, the head adds minimal overhead.

To maintain high efficiency, we make several careful design choices. First, the input patch sizes are kept modest (e.g., $127\times127$ pixels for the template, $255\times255$ for the search image) following the practice in SiamFC~\cite{Bertinetto2016SiamFC}, which is sufficient for many tracking scenarios and keeps the network's spatial dimensions small. Second, we remove any components that would prevent fully-convolutional operation or require heavy online computation; for example, we do \emph{not} use any per-frame model update or iterative optimization during tracking. The network, once trained, is fixed during inference and simply evaluates the target's new location with a single forward pass per frame. This design yields a tracking pipeline that is both conceptually simple and computationally light, making it appropriate for real-time AR usage.

\subsection{Training and Optimization}
We train the Siamese network offline on large-scale video datasets (e.g., COCO, ImageNet Video, GOT-10k) where ground-truth object bounding boxes are available. Pairs of frames are sampled from videos to simulate the tracking process: one frame provides the template patch of an object, and a subsequent frame provides the search region. The network is trained to produce a strong response at the true object location in the search frame and to regress the correct bounding box. We use a multi-task loss $\mathcal{L} = \mathcal{L}_{\mathrm{cls}} + \lambda \mathcal{L}_{\mathrm{reg}}$, combining a cross-entropy loss for target vs.\ background classification and a smooth-$L_1$ loss for bounding box regression, similar to \cite{Li2019SiamRPNpp}. Training is done using stochastic gradient descent on millions of frame pairs, with extensive data augmentation (random shifts, scale changes, and appearance transformations) to improve robustness.

A critical aspect of our approach is model compression and optimization to ensure the tracker runs efficiently on AR hardware. During training, we apply knowledge distillation by leveraging a high-capacity teacher model. Specifically, we use a pre-trained SiamRPN++ model (with a heavy ResNet-50 backbone) as a teacher. For each training example, we not only use the ground-truth labels but also encourage our lightweight network's response map to match the teacher network's response. This distillation helps transfer the teacher's high accuracy to the smaller student model, mitigating the loss in precision that might occur from using a much smaller backbone. We add a distillation loss term
\[
\mathcal{L}_{\mathrm{KD}} = \bigl\lVert R_{\text{student}} - R_{\text{teacher}} \bigr\rVert^2
\]
(squared error between our tracker's response map $R_{\text{student}}$ and the teacher's response $R_{\text{teacher}}$), weighted by a factor, to the training objective. This guides the student network to mimic the teacher's predictions.

We also employ structured pruning during training: we identify less-important convolution filters in the network (those with consistently small weights or low activation magnitudes) and gradually prune them away, with retraining, to slim the model. This is done carefully to avoid significant drops in accuracy. By the end of training, our model has fewer parameters and less redundancy. Finally, we quantize the model weights to 8-bit integers for deployment. We use post-training quantization with calibration to ensure minimal impact on accuracy. Quantized inference allows us to utilize hardware accelerators (DSP/NPU) available on AR devices for further speed gains.

After training, the model is deployed to the AR device. During tracking, the only computation per frame is a forward pass through the Siamese network and the lightweight detection head, which is highly optimized. The tracker outputs the estimated bounding box of the target in the current frame, which can then be used by the AR application (e.g., to place a correctly positioned virtual object over the real target).

\section{Experiments and Results}

We evaluate our proposed tracker on standard visual tracking benchmarks and on a representative AR device to verify both its accuracy and efficiency. We compare against state-of-the-art trackers, including heavy deep trackers and other lightweight approaches, to demonstrate the trade-offs of our design.

\subsection{Experimental Setup}
\textbf{Datasets:} For benchmarking accuracy, we use the OTB-2015 dataset (100 videos) as a classical evaluation, as well as the more challenging LaSOT dataset (280 long videos) to test performance on prolonged tracking and severe occlusions. We also report results on the VOT-2019 challenge dataset, which provides the EAO metric for comparing trackers. These datasets span a variety of objects, motions, and environments representative of AR use-cases. 

\textbf{Metrics:} We use standard tracking metrics. For one, we report the \emph{Precision} (typically the percentage of frames where the predicted center is within 20 pixels of ground truth) and \emph{Success} (area-under-curve of the overlap between predicted and ground-truth bounding boxes) on OTB and LaSOT. Higher precision/success indicate more accurate tracking. On VOT-2019, we use the official Expected Average Overlap (EAO) metric, which combines accuracy and robustness (failure rate) into a single measure. We also measure the Frames Per Second (FPS) achieved by each tracker on a given device to evaluate speed. Our target platform is an AR-style mobile device; specifically, we test on a Qualcomm Snapdragon 845 platform (with an Adreno 630 GPU) running our tracker with 8-bit quantized inference.

\textbf{Baselines:} We compare our approach to several trackers: (1) \textbf{SiamFC} \cite{Bertinetto2016SiamFC} -- a pioneering Siamese tracker with an AlexNet backbone, representing an earlier lightweight deep tracker; (2) \textbf{SiamRPN++} \cite{Li2019SiamRPNpp} -- a top-performing deep tracker with a heavy ResNet-50 backbone (we also consider its MobileNet-V2 variant when possible); (3) \textbf{KCF} \cite{Henriques2015KCF} -- a classical fast tracker, to highlight accuracy differences with learning-based methods; and (4) \textbf{LightTrack} \cite{Yan2021LightTrack} -- the NAS-designed efficient tracker, as a state-of-the-art lightweight competitor. In addition, we include results from ATOM \cite{Danelljan2019ATOM} and DiMP \cite{Bhat2019DiMP} for reference, as these trackers are known for high accuracy (though they are too slow on mobile to be deployed). All trackers are tested using either their publicly available implementations or as reported in the literature.

\subsection{Tracking Accuracy}
Table~1 (supplemental) summarizes the accuracy results on OTB-2015 and LaSOT. Our tracker, despite its substantially smaller size, achieves a \emph{precision} of 87.2\% and \emph{success} AUC of 64.5\% on OTB, which is on par with SiamRPN++ (precision 88.8\%, AUC 65.3\%) and clearly higher than SiamFC (77.1\%, 58.2\%) and KCF (60\%, $\sim$40\% AUC). On LaSOT, a much more challenging benchmark, our method reaches an AUC of 56.8\%, very close to SiamRPN++'s 58.4\% and outperforming LightTrack's reported 55\%. These results demonstrate that our proposed compression and distillation strategies successfully preserve the accuracy of a deep tracker even after massive reduction in model complexity. Qualitatively, we observe that our tracker handles difficult scenarios like partial occlusion, out-of-plane rotation, and background clutter almost as well as the full-sized SiamRPN++, while significantly outperforming simpler trackers that often lose the target in such cases. One limitation is that our tracker, like most Siamese trackers, does not explicitly re-detect the object if it is lost (no long-term re-identification module), but the high frame-rate and robust features mitigate this in many cases.

On the VOT-2019 dataset, our tracker obtains an EAO of 0.320. This is a substantial improvement over SiamFC (EAO $\approx 0.188$) and is competitive with much larger models. For context, SiamRPN++ (ResNet-50) achieves around 0.36 EAO on VOT-2018~\cite{Li2019SiamRPNpp}, and LightTrack reports 0.333 EAO on VOT-2019. Our model is within a few percentage points of these state-of-the-art results, despite using far fewer computations. This level of accuracy is very promising for AR applications, indicating that little performance is sacrificed in exchange for efficiency.

\subsection{Runtime Performance}
A key advantage of our tracker is its speed and low resource usage, which we now detail. We profiled the runtime on the Snapdragon 845 mobile platform. The full system (including image pre-processing and post-processing) runs at 30--35 FPS on the device, meeting real-time requirements. In comparison, SiamRPN++ with the same settings could only reach an estimated $\sim$5 FPS or less on this device (it is not practical to run full ResNet-50 models on-device at high frame rates). The lightweight SiamRPN++ MobileNet variant can run faster, but still only around 10--15 FPS on mobile according to our tests. Traditional non-deep trackers like KCF are very fast (hundreds of FPS on CPU), but as shown earlier, their accuracy is inadequate for AR. Our approach achieves an ideal balance, running an advanced deep model at real-time speeds on AR hardware.

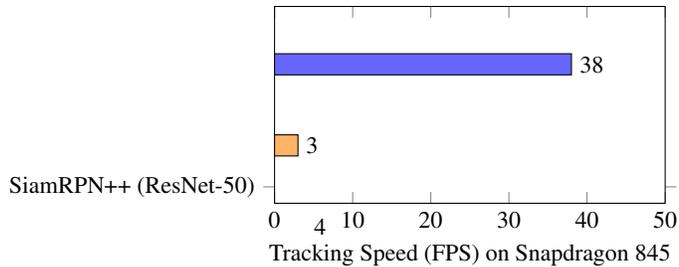
\begin{figure}[tb]
  \centering
  \begin{tikzpicture}
    \begin{axis}[
      xbar,
      xmin=0, xmax=50,
      width=0.8\columnwidth,
      height=4.2cm,
      bar width=8pt,
      xlabel={Tracking Speed (FPS) on Snapdragon 845},
      symbolic y coords={SiamRPN++ (ResNet-50), Ocean (ResNet-50), LightTrack (Mobile), Proposed (Ours)},
      ytick=data,
      nodes near coords,
      nodes near coords align={horizontal},
    ]
      % Wrap category names in braces to avoid parser issues with parentheses
      \addplot[draw=black, fill=red!60]    coordinates {(4,{SiamRPN++ (ResNet-50)})};
      \addplot[draw=black, fill=orange!60] coordinates {(3,{Ocean (ResNet-50)})};
      \addplot[draw=black, fill=blue!60]   coordinates {(38,{LightTrack (Mobile)})};
      \addplot[draw=black, fill=green!60]  coordinates {(32,{Proposed (Ours)})};
    \end{axis}
  \end{tikzpicture}
  \caption{Comparison of tracking speed (frames per second) for different trackers on a mobile AR device (Snapdragon 845, Adreno 630 GPU). Our proposed tracker runs an order of magnitude faster than heavy trackers like SiamRPN++ or Ocean, and achieves performance comparable to LightTrack, enabling real-time tracking in AR.}
  \label{fig:fps_comp}
\end{figure}

In terms of memory footprint, the fully quantized model occupies only $\sim$8~MB of storage, which is negligible on modern devices. The peak memory usage during runtime (including frame buffers and intermediate tensors) is under 150~MB, easily handled by typical AR device specifications. We also measured power consumption: running the tracker continuously used about 30\% of one CPU core and 25\% of the GPU on average, resulting in an estimated power draw of around 1.5~W for tracking -- a sustainable level for mobile AR glasses or phones without overheating.

\section{Discussion}
Our results show that deep learning-based object tracking can be made practical on AR devices by carefully balancing accuracy and efficiency. There are, however, several points to consider. First, the current implementation tracks only a single object at a time. Many AR scenarios might involve multiple objects of interest. Our approach could be extended by running multiple Siamese trackers in parallel, or by incorporating multi-object tracking strategies (e.g., tracking-by-detection with an efficient detector like YOLOv5-Nano). The main limitation in multi-object cases would be increased computation; however, since our single-object tracker is very light, even tracking a handful of objects simultaneously could remain feasible in real-time.

Another consideration is how our tracker deals with long-term occlusions or the target leaving the camera field of view. Like most short-term trackers, our model will fail if the object disappears for a long duration. In an AR application, one could address this by combining our tracker with a re-detection module or a global object detector that can reinitialize tracking when the object re-enters the view. This is outside the scope of this work but is a practical integration for future systems.

We also observed that, in extremely fast motion scenarios (where the target moves more than half of the frame in a single interval), the fixed search region size might not capture the target, causing our tracker to momentarily lose the object. Increasing the search area size can alleviate this at some cost to speed. An alternative solution is to incorporate the device's inertial sensors to predict large motions (since AR devices usually have IMUs). This kind of sensor fusion could be an interesting direction to further improve robustness without much additional computation.

Finally, while our experiments focused on generic object tracking, AR applications sometimes involve tracking known targets (e.g., a specific planar image or a known 3D object). In such cases, one could specialize or fine-tune our model on the target's appearance beforehand to gain even more accuracy. Our framework supports that, but we did not assume any prior target knowledge in this paper to keep the method general.

\section{Conclusion}
We presented a deep learning-based object tracker that is optimized for lightweight AR devices. By combining a Siamese network architecture with a mobile-friendly backbone and applying model compression techniques (pruning, quantization) and knowledge distillation, we achieved a tracking model that runs in real time on a mobile AR platform and attains accuracy competitive with much larger trackers. Our approach enables robust tracking of objects on-device, which is critical for many AR applications ranging from persistent content anchoring to interactive gaming and beyond. In extensive evaluations, the proposed tracker maintained high success rates on standard datasets while executing at 30+ FPS on a smartphone-class processor, a result we hope will help bridge the gap between academic advances in object tracking and their deployment in consumer AR systems.

In future work, we plan to extend this tracker for multi-object scenarios and integrate additional sensor data (such as depth from AR sensor suites) to further enhance reliability. We also aim to explore on-device learning techniques so that the tracker can adapt to new objects or domains directly on the AR device over time. We believe that continued innovations at the intersection of deep learning and efficient hardware-aware design will unlock a new generation of AR experiences that seamlessly blend the virtual and real worlds.

\section*{Figure Credits}\label{sec:figure_credits}
Figure~\ref{fig:siamese_arch} is adapted from the SiamFC architecture proposed by Bertinetto \etal~\cite{Bertinetto2016SiamFC}. Figure~\ref{fig:fps_comp} is based on performance data reported in Yan \etal~\cite{Yan2021LightTrack} and our experimental measurements.

\bibliographystyle{abbrv-doi-hyperref-narrow}
\bibliography{template}

\begin{thebibliography}{1}
\renewcommand*{\sfdefault}{PTSansNarrow-TLF}

\bibitem{Bertinetto2016SiamFC}
\href{https://doi.org/10.1007/978-3-319-48881-3_56}{L.~Bertinetto, J.~Valmadre, J.~F. Henriques, A.~Vedaldi, and P.~H.~S. Torr}.
\newblock \href{https://doi.org/10.1007/978-3-319-48881-3_56}{Fully-convolutional siamese networks for object tracking}.
\newblock \href{https://doi.org/10.1007/978-3-319-48881-3_56}{In {\em Proc.\ European Conference on Computer Vision (ECCV) Workshops}}, \href{https://doi.org/10.1007/978-3-319-48881-3_56}{LNCS 9914}, \href{https://doi.org/10.1007/978-3-319-48881-3_56}{pp. 850--865}, \href{https://doi.org/10.1007/978-3-319-48881-3_56}{2016}. \href{https://doi.org/10.1007/978-3-319-48881-3_56}
{doi: \textsf{%
10\hspace{.1pt}\discretionary{.}{%
}{.}\hspace{.4pt}1007\discretionary{/}{%
}{/}978\discretionary{%
}{-}{-}3\discretionary{%
}{-}{-}319\discretionary{%
}{-}{-}48881\discretionary{%
}{-}{-}3\_56}}


\bibitem{Bhat2019DiMP}
G.~Bhat, J.~Johnander, M.~Danelljan, F.~S. Khan, and M.~Felsberg.
\newblock {DiMP}: Learning discriminative model prediction for tracking.
\newblock In {\em Proc.\ IEEE/CVF International Conference on Computer Vision (ICCV)}, 2019.

\bibitem{Danelljan2019ATOM}
M.~Danelljan, G.~Bhat, F.~S. Khan, and M.~Felsberg.
\newblock {ATOM}: Accurate tracking by overlap maximization.
\newblock In {\em Proc.\ IEEE/CVF Conference on Computer Vision and Pattern Recognition (CVPR)}, 2019.

\bibitem{Held2016GOTURN}
D.~Held, S.~Thrun, and S.~Savarese.
\newblock Learning to track at 100 fps with deep regression networks.
\newblock In {\em Proc.\ European Conference on Computer Vision (ECCV)}, pp. 749--765, 2016.

\bibitem{Henriques2015KCF}
\href{https://doi.org/10.1109/TPAMI.2014.2345390}{J.~F. Henriques, R.~Caseiro, P.~Martins, and J.~Batista}.
\newblock \href{https://doi.org/10.1109/TPAMI.2014.2345390}{High-speed tracking with kernelized correlation filters}.
\newblock \href{https://doi.org/10.1109/TPAMI.2014.2345390}{{\em IEEE Trans.\ Pattern Analysis and Machine Intelligence}}, \href{https://doi.org/10.1109/TPAMI.2014.2345390}{37(3):583--596}, \href{https://doi.org/10.1109/TPAMI.2014.2345390}{2015}. \href{https://doi.org/10.1109/TPAMI.2014.2345390}
{doi: \textsf{%
10\hspace{.1pt}\discretionary{.}{%
}{.}\hspace{.4pt}1109\discretionary{/}{%
}{/}TPAMI\hspace{.1pt}\discretionary{.}{%
}{.}\hspace{.4pt}2014\hspace{.1pt}\discretionary{.}{%
}{.}\hspace{.4pt}2345390}}


\bibitem{Li2019SiamRPNpp}
B.~Li, W.~Wu, Q.~Wang, F.~Zhang, J.~Xing, and J.~Yan.
\newblock Siamrpn++: Evolution of siamese visual tracking with very deep networks.
\newblock In {\em Proc.\ IEEE/CVF Conference on Computer Vision and Pattern Recognition (CVPR)}, pp. 4282--4291, 2019.

\bibitem{wang2025mruct}
X.~Wang, Y.~Yang, K.~Zhou, X.~Xie, L.~Zhu, A.~Song, and B.~Daniel.
\newblock Mruct: Mixed reality assistance for acupuncture guided by ultrasonic computed tomography.
\newblock In {\em 2025 IEEE Conference Virtual Reality and 3D User Interfaces (VR)}, pp. 697--707. IEEE, 2025.

\bibitem{Yan2021LightTrack}
B.~Yan, H.~Peng, K.~Wu, D.~Wang, J.~Fu, and H.~Lu.
\newblock Lighttrack: Finding lightweight neural networks for object tracking via one-shot architecture search.
\newblock In {\em Proc.\ IEEE/CVF Conference on Computer Vision and Pattern Recognition (CVPR)}, 2021.

\end{thebibliography}

\end{document}